\documentclass[prb, twocolumn, aps,amsmath,amssymb,amsfonts, superscriptaddress]{revtex4}
\usepackage[german,american,english]{babel}
\usepackage{graphicx}% Include figure files
\usepackage{graphics}% Include figure files
\usepackage{dcolumn}% Align table columns on decimal point
\usepackage{bm}% bold math
\usepackage{amssymb}
\usepackage{amsmath}
\usepackage{amsfonts}
\usepackage{epsfig}

\newcommand {\bkt} [1] {\langle #1 \rangle}
\newcommand {\dbkt} [2] {\langle #1 | #2 \rangle}
\newcommand {\tbkt} [3] {\langle #1 | #2 | #3 \rangle}

 \newcommand {\beq}{\begin{equation}}
\newcommand {\eeq}{\end{equation}}
\begin{document}
\title{Realizing singlet-triplet qubits in multivalley Si quantum dots}
\author{Dimitrie Culcer}
\affiliation{Condensed Matter Theory Center, Department of Physics, University of Maryland, College Park MD20742-4111}
\author{\L ukasz Cywi\'nski}
\affiliation{Condensed Matter Theory Center, Department of Physics, University of Maryland, College Park MD20742-4111}
\author{Qiuzi Li}
\affiliation{Condensed Matter Theory Center, Department of Physics, University of Maryland, College Park MD20742-4111}
\author{Xuedong Hu}
\affiliation{Joint Quantum Institute, Department of Physics, University of Maryland, College Park MD20742-4111}
\affiliation{Department of Physics, University at Buffalo, SUNY, Buffalo, NY 14260-1500}
\author{S.~Das Sarma}
\affiliation{Condensed Matter Theory Center, Department of Physics, University of Maryland, College Park MD20742-4111}
\begin{abstract}
There has been significant progress in the implementation and manipulation of singlet-triplet qubits in GaAs quantum dots. Given the considerably longer spin coherence times measured in Si, considerable interest has been generated recently in Si quantum dots. The physics of these systems is considerably more complex than the physics of GaAs quantum dots owing to the presence of the valley degree of freedom, which constitutes the focus of this work. In this paper we investigate the physics of Si quantum dots and focus on the feasibility of quantum coherent singlet-triplet qubit experiments analogous to those performed in GaAs. This additional degree of freedom greatly increases the complexity of the ground state and gives rise to highly nontrivial and interesting physics in the processes of qubit initialization, coherent manipulation and readout. We discuss the operational definition of a qubit in Si-based quantum dots. We find that in the presence of valley degeneracy a singlet-triplet qubit cannot be constructed, whereas for large valley splitting ($\gg k_BT$) the experiment is similar to GaAs. We show that experiments on singlet-triplet qubits analogous to those in GaAs would provide a method for estimating the valley coupling in Si. A Zeeman field distinguishes between different initialized states for \textit{any} valley splitting and provides a tool to determine the size of this splitting.
\end{abstract} 
\maketitle

\section{Introduction}

Spin-based qubits are seen as promising candidates for scalable quantum computation, with donor \cite{Kane} and quantum-dot \cite{LDV} spins at the focus of research. Electrical readout and control of single spins in quantum dots (QDs) have proven challenging, yet GaAs double quantum dots (DQDs), where spin blockade \cite{Ono} and charge sensors \cite{PettaSci} enable observation of single/two-spin dynamics \cite{PettaSci, Koppens, Taylor, PettaRMP}, have seen impressive experimental progress.  In this article we establish the precise criteria for realizing spin qubits in Si QDs, where the multivalley structure of the ground state introduces fundamental complications in distinguishing spin and orbital degrees of freedom.

The original proposal by Loss and DiVincenzo \cite{LDV} made use of a quantum dot array, in which one electron spin on each dot constitutes the qubit. More recently there has been significant progress in implementing an alternative scheme,\cite{PettaSci} in which the singlet and triplet states of two electrons in a DQD make up the qubit.\cite{Jacak, Levy, Mohseni, Taylora, Taylorb, Barthel} One particular successful experiment involves initialization, manipulation, and measurement of two-spin singlet and triplet states \cite{PettaSci}.  Here a (0,2) singlet state is initialized, where ($n$,$m$) indicates the occupancy of the left and right dots. Since the (0,2)---single dot---singlet and triplet are separated by an meV gap, initialization of the singlet is easy and reliable. Tuning the gate voltages then allows tunneling of one electron to the left dot to form a (1,1) singlet.  When the bias is pushed deep into the (1,1) regime [where (1,1) is by far the electrostatic ground state configuration], the singlet and triplet are essentially degenerate due to the small tunnel coupling between the dots, so that a small magnetic field inhomogeneity (e.g. due to the Overhauser field of the nuclei) between the dots can rotate the two-electron states between the singlet and the triplet. After some mixing time in the (1,1) regime, tuning the bias returns the system to the (0,2) configuration, where electrical readout is possible due to spin blockade \cite{PettaRMP}.  This experiment clearly illustrates the existence of quantum coherence in the DQD system, and the distinct possibility of using the two-electron singlet and unpolarized triplet as the two states of a logical qubit, with reliable initialization, single-qubit rotation, and measurement. 

Silicon is often regarded as the best semiconducting host material for spin qubits because of its excellent spin coherence properties: spin-orbit coupling is very small, the hyperfine interaction can be reduced by isotopic purification \cite{Witzel}, and the electron-phonon interaction is weak as well. Furthermore, the mature Si microfabrication technology will help attempts to scale up a Si-based quantum computer (QC) architecture.  At present, Si/SiGe \cite{Copper} and Si/SiO$_2$ \cite{Nordberg} quantum dots, and Si:P \cite{Stegner} are being actively investigated and progress has been made in spin blockade in Si quantum dots.\cite{Liu} The biggest obstacle to spin QC in Si is valley degeneracy: bulk Si has six degenerate conduction band minima.  While this degeneracy can be reduced by strain or the presence of an interface, it complicates the orbital and spin state spectrum \cite{Hada} and leads to valley-interference effects for spin interactions \cite{Koiller}. At the Si/SiO$_2$ interface only two valleys are relevant to the ground orbital state. Scattering at the interface further lifts the valley degeneracy by producing a valley-orbit coupling $\Delta$. The magnitude of $\Delta$ is generally not known {\it a priori} and is sample-dependent \cite{Friesen}.  Currently measurement of valley splitting $\Delta$ is generally done for 2D electron gases at high magnetic fields, and the zero-field valley splitting is extrapolated \cite{Private}. We note that the case of large valley splitting has been examined in a number of recent publications.\cite{Qiuzi, Erik} 

In this article we study the physics of Si-based quantum dots and the feasibility of experiments analogous to Ref.~[\onlinecite{PettaSci}] in a Si/SiO$_2$ (or Si/SiGe \cite{Copper}) DQD, focusing on the effects of the valley degree of freedom on qubit initialization, operation, and spin blockade within the effective mass approximation. We identify the conditions required for an operational singlet-triplet qubit in Si. We further demonstrate that a quantum coherent experiment analogous to Ref.~[\onlinecite{PettaSci}] may provide a direct way to estimate the valley splitting $\Delta$. While our discussion focuses on Si/SiO$_2$ and is directly relevant to experiments on Si/SiO$_2$ quantum dots \cite{Private}, the findings are generally applicable to Si quantum dots. In addition, we expect our findings to be at least qualitatively relevant to other systems in which the valley degree of freedom plays an important role, such as carbon, in which significant progress has been made lately.\cite{Churchill, Palyi}

The outline of this paper is as follows. We introduce the model of the DQD in Section \ref{sec:DQD}. We proceed to study the initialization process in singlet-triplet qubits n Section \ref{sec:init} followed by manipulation of the qubit in Section \ref{sec:manip}. In Section \ref{sec:msr} we demonstrate that a quantum coherent experiment on singlet-triplet qubits can be used to estimate the valley splitting. Section \ref{sec:issues} is devoted issues specific to silicon, such as interface roughness and the need for an external inhomogeneous magnetic field. Finally, Section \ref{sec:summary} contains a summary of our findings.

\section{Double quantum dot}
\label{sec:DQD}

We choose nominally $\hat{\bm z}$ as the growth direction for the Si/SiO$_2$ heterostructure we consider.  The two dots are located at ${\bm R}_{R,L} = (\pm X_0, 0, 0)$, where $R$ and $L$ stand for right and left respectively. The Hamiltonian is $H=H_0 + H_v$, with $H_0 = \big(\sum_{i=1,2}T^{(i)} + V_Q^{(i)}\big) + V_{ee}$, where $T$ is the kinetic energy operator and $V_Q$ the confinement potential 
\begin{equation}
\arraycolsep 0.3 ex
\begin{array}{rl}
\displaystyle V_Q = & \displaystyle (1/2) \, m_t \omega_0^2 \, {\rm Min} [(x-X_0)^2, (x+X_0)^2] - eEx \\ [1ex]
+ & \displaystyle (1/2) \, m_t \omega_0^2 y^2 + (1/2) \, m_z \omega_z^2 z^2,
\end{array} 
\end{equation}
with $m_t$ and $m_z$ respectively the in- and out-of-plane Si effective masses. The Coulomb interaction between electrons at ${\bm r}_1$ and ${\bm r}_2$ is $V_{ee} = e^2/(\epsilon|{\bm r}_1 - {\bm r}_2|)$, where $\epsilon = (\epsilon_{Si} + \epsilon_{SiO_2})/2$ includes the image charge in the SiO$_2$ layer. $H_v$ is a single-particle phenomenological coupling between the valleys discussed below. The electric field $E$ raises the energy of the left dot with respect to the right dot. The confinement potential and ground state for $E=0$ are identical in each dot, with the single-dot potentials $V_{R, L}(x) = (1/2) \, m_t \omega_0^2 \, (x \mp X_0)^2$.  At the Si/SiO$_2$ interface the lowest valleys are at $\pm \bkt{k_z}$, with $\bkt{k_z} = 0.85 (2\pi/a_{Si})$, and the lattice constant $a_{Si} \! = 5.43 {\rm\AA}$.  The ground-state single-electron wave functions $R_{z, \bar{z}}$ and $L_{z, \bar{z}}$ represent the degenerate $\pm \bkt{k_z}$ valleys on the right and left dots respectively.  In the right dot $(T + V_R) \, R_{z, \bar{z}} = \varepsilon_0 \, R_{z, \bar{z}}$, with $R_{z, \bar{z}} = F_R ({\bm r} - {\bm R}_R) e^{\pm i{\bm k}_z \cdot({\bm r} - {\bm R}_R)} u_{z, \bar{z}} ({\bm r} - {\bm R}_R)$, and on the left $R \rightarrow L$.  The envelope functions are
\begin{equation}
F_{R, L}({\bm r} - {\bm R}_{R, L}) = \frac{1}{\pi^{3/4} (a^2b)^{1/2}} \, e^{-\frac{(x \mp X_0)^2}{2a^2}} e^{-\frac{y^2}{2a^2}}e^{-\frac{z^2}{2b^2}},
\end{equation}
where $a = \sqrt{\frac{\hbar}{m_t\omega_0}}$ and $b = \sqrt{\frac{\hbar}{m_z\omega_z}}$ the in-plane (Fock-Darwin radius) and growth-direction confinement length. The lattice-periodic Bloch function $ u_{z, \bar{z}}({\bm r}) = \sum_{\bm K} c^{z, \bar{z}}_{\bm K} e^{i{\bm K}\cdot{\bm r}}$ with ${\bm K}$ reciprocal lattice vectors. The overlap $\dbkt{L_{z, \bar{z}}}{R_{z, \bar{z}}} = e^{-d^2}$ where $d = X_0/a$.  Overlaps such as $\dbkt{L_z}{L_{\bar{z}}}, \dbkt{L_{z}}{R_{\bar{z}}}$ are suppressed by an exponential of the form $e^{-\frac{b^2Q_z^2}{4}}$, where $Q_z = \frac{2\pi n_z}{a_{Si}} - 2\bkt{k_z}$, with $n_z$ an integer. Such an exponential appears in all but one of the matrix elements of $H_0$ involving functions from different valleys. All such intervalley terms can be neglected except one, discussed below. The only nonzero matrix elements of $H_v$ are $\tbkt{L_{z, \bar{z}}}{H_v}{L_{\bar{z}, z}} = \tbkt{R_{z, \bar{z}}}{H_v}{R_{\bar{z}, z}} = \Delta$, with $\Delta\!  >\! 0$ and assuming $\Delta$ has the same form on each dot. We define also $\varepsilon_R = \tbkt{R_{z, \bar{z}}}{(T+V_Q)}{R_{z, \bar{z}}}$, $\varepsilon_L = \tbkt{L_{z, \bar{z}}}{(T+V_Q)}{L_{z, \bar{z}}}$, and the dimensionless detuning as $(\varepsilon_L - \varepsilon_R)/(2d\varepsilon_0)$. Diagonalizing the single-particle Hamiltonian with the valley coupling we obtain the \textit{valley eigenstates} $R_\pm = (1/\sqrt{2}) \, (R_z \pm R_{\bar {z}})$ and $L_\pm = (1/\sqrt{2}) \, (L_z \pm L_{\bar {z}})$ with corresponding energies $\varepsilon_{R,L} \pm \Delta$. We orthogonalize these following Ref.\ [\onlinecite{Burkard}], with $\tilde{L}_\pm = (L_\pm - gR_\pm)/\sqrt{1 - 2lg + g^2}$, with $g = (1- \sqrt{1 - l^2})/l$; for $\tilde{R}_\pm$ one swaps $L \leftrightarrow R$. These are the states that will be used henceforth.

\begin{figure}[tbp] 
\begin{tabular}{lr}
(a) \,\,\,\,\,\,\,\,\,\,\,\,\,\,\,\,\,  \,\,\,\,\,\,\,\,\,\,\,\,\,\,\,\,\,  \,\,\,\,\,\,\,\,\,\,\,\,\,\,\,\,\,  \,\,\,\,\,\,\,\,\,\,\,\,\, (b) \\
\includegraphics[width=0.75\columnwidth]{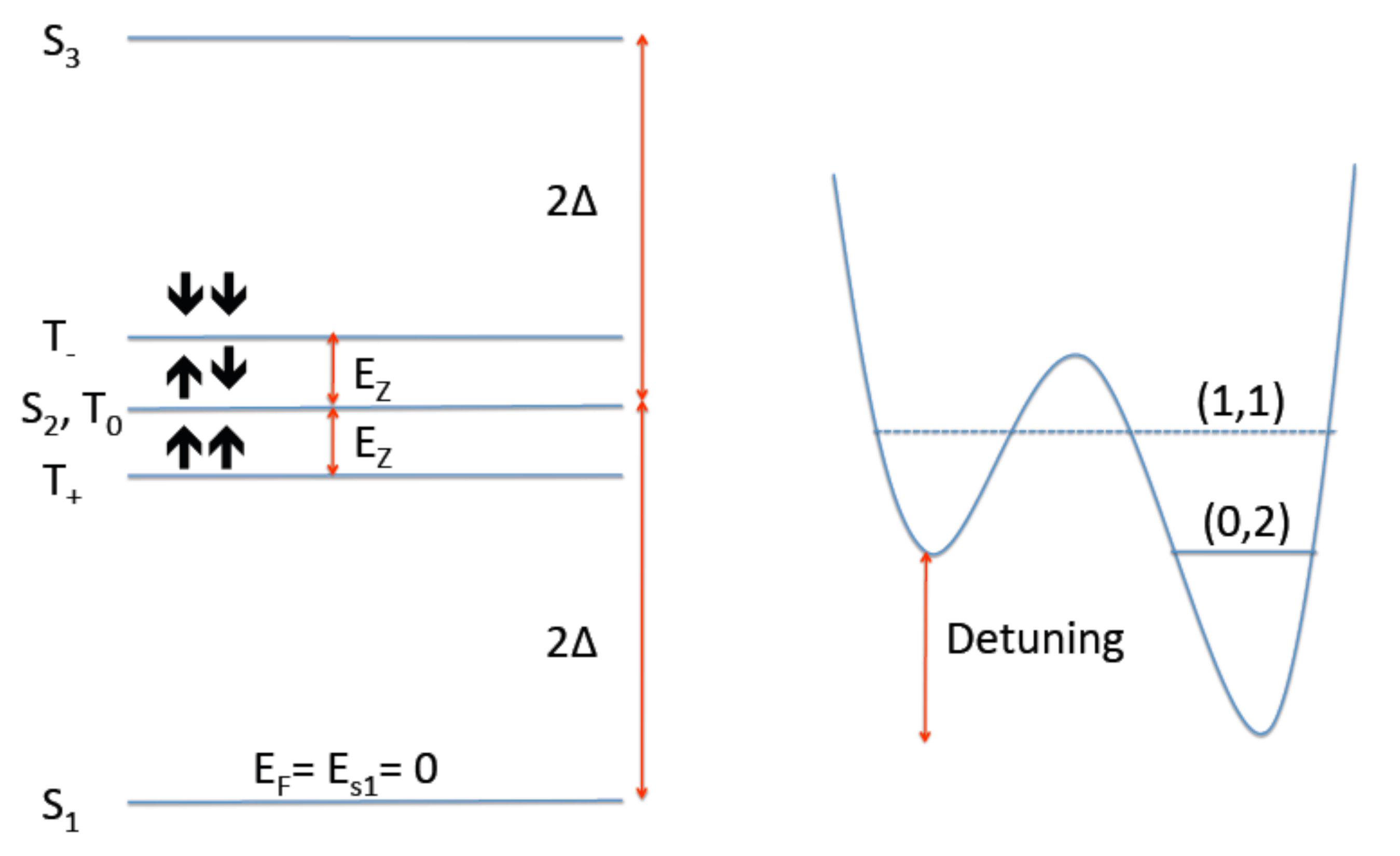}
\end{tabular} 
\caption{(a) Single dot energy levels for finite $\Delta$ and ${\bm B}$ such that $2\Delta>E_Z$. The lowest energy state is the singlet $S_1$, followed by the triplet $T_+$, the degenerate singlet $S_2$/triplet $T_0$ and triplet $T_-$, and the singlet $S_3$. Spin orientations of electrons in triplet states are indicated by arrows. (b) Schematic of the biased double dot. During initialization the detuning is large, and the (0,2) states are lowest in energy. After loading the detuning is lowered and the (1,1) states are at the same energy as the (0,2) states -- the charge transition regime. An inhomogeneous magnetic field mixes the singlets and triplets. }
\label{Schematic}
\end{figure}

\section{Initialization}
\label{sec:init}
We begin by studying the initialization process, which involves loading two electrons onto the right dot. For this purpose it is imperative to analyze first the spectrum of the doubly-occupied right dot, that is the configuration (0,2). The lowest-energy two-particle spatial wave functions are $\displaystyle \phi_{S1,S3} = \tilde{R}_\mp^{(1)} \tilde{R}_\mp^{(2)}$ and
\begin{equation}\label{SingleDot}
\begin{array}{rl}
\displaystyle \phi_{S2} = & \displaystyle \frac{1}{\sqrt{2}}\, \big(\tilde{R}_+^{(1)} \tilde{R}_-^{(2)} + \tilde{R}_+^{(2)} \tilde{R}_-^{(1)}\big) \\ [3ex]
\displaystyle \phi_T = & \displaystyle \frac{1}{\sqrt{2}}\, \big(\tilde{R}_+^{(1)} \tilde{R}_-^{(2)} - \tilde{R}_+^{(2)} \tilde{R}_-^{(1)}\big), 
\end{array}
\end{equation}
where $(i)$ denotes the $i$th electron.  In the basis $\{ \phi_{S1}, \phi_{S2} , \phi_T, \phi_{S3}\}$ the matrix elements of the Hamiltonian are $2\varepsilon_R + u + {\rm diag} \, (-2\Delta, 0, 0, 2\Delta)$, where $u = \int d^3 r_1 \int d^3 r_2\, R_{z, \bar{z}}^{*(1)} R_{z, \bar{z}}^{*(2)} V_{ee} \, R_{z, \bar{z}}^{(1)} R_{z, \bar{z}}^{(2)}$. The valley-exchange Coulomb integral $\int d^3 r_1 \int d^3 r_2\, R_z^{*(1)} R_{\bar{z}}^{*(2)} V_{ee} \, R_{\bar{z}}^{(1)} R_z^{(2)}$ is not suppressed by an exponential, however we find its value to be $\ll 1\mu$eV and it will therefore be assumed of no consequence henceforth. For $\Delta\!=\!0$ and no external magnetic field ${\bm B}$, all levels are degenerate so that it is impossible to load any particular two-electron state. The spectrum for finite $\Delta$ and ${\bm B}$, yielding a Zeeman energy $E_Z$ with $2\Delta > E_Z$, is shown in Fig.~\ref{Schematic}.  The triplet states thus split into $T_+$, $T_0$, and $T_-$, separated in energy by $E_Z$. 

The loading process makes use of an outside reservoir with Fermi energy $\varepsilon_F$ which is thermally broadened by $\sim$ k$_B$T. The reservoir is tuned to be on resonance with the lowest-energy singlet state. Here we neglect the the differences in tunnel couplings between the various states and the reservoir. Consequently the probability of loading any of the states is proportional to the Fermi distribution at its energy. If $\Delta \gg$ k$_B$T, the lowest-energy singlet state can be loaded exclusively. Numerically $\Delta \approx$ 0.1meV is sufficient to fulfil this condition at dilution refrigerator temperatures of T=100mK, where k$_B$T $\approx$ 0.01meV.  In this regime the two-electron initialization process in a Si DQD is identical to the GaAs DQD in Ref.~[\onlinecite{PettaSci}].

\section{Qubit manipulation}
\label{sec:manip}
Manipulation of the singlet-triplet qubit involves switching to the configuration (1,1), where singlet-triplet mixing is achieved through the us of an inhomogeneous ${\bm B}$. We study the Hilbert space of two electrons in a Si DQD.  We do not include the high-energy (2,0) states.  The seven space-symmetric Hund-Mulliken (HM) wave functions of the lowest-energy singlets are $\phi_{S1}$, $\phi_{S2}$, $\phi_{S3}$ and the functions
\begin{equation}\label{Singlets}
\arraycolsep 0.3 ex
\begin{array}{rl}
\displaystyle \phi_S^{\pm\pm} = & \displaystyle \frac{1}{\sqrt{2}} \, (\tilde{L}_\pm^{(1)} \tilde{R}_\pm^{(2)} + \tilde{L}_\pm^{(2)} \tilde{R}_\pm^{(1)}) \\ [3ex]
\displaystyle \phi_S^{m\pm} = & \displaystyle \frac{1}{\sqrt{2}}\, (\tilde{L}_\pm^{(1)} \tilde{R}_\mp^{(2)}  + \tilde{L}_\pm^{(2)} \tilde{R}_\mp^{(1)}).
\end{array}
\end{equation}
These singlet states split into three uncoupled subspaces. The $ \{\phi_S^{++}, \phi_{S3} \}$ and $\{\phi_S^{--}, \phi_{S1} \}$ subspaces are composed of HM wave functions where the two electrons are in the same valley eigenstate, while the subspace $\{\phi_S^{m+}, \phi_S^{m-}, \phi_{S2} \}$ consists of wave functions where electrons are in different valley eigenstates. The five antisymmetric counterparts of the states in Eq.~(\ref{Singlets}), denoted by $\phi_T^{\pm\pm}$, $\phi_T^{m\pm}$ and $\phi_{T2}$, are evident (clearly $\phi_{S1}$ and $\phi_{S3}$ do not have antisymmetric counterparts.) These triplets in turn split into three subspaces, with $\{\phi_T^{++}\}$ and $ \{\phi_T^{--}\}$ single-valley HM triplets, and $\{\phi_T^{m+}, \phi_T^{m-}, \phi_T^{md} \}$ mixed-valley triplet states.  Since the overlap between states from different valleys is negligible, matrix elements of the form $\tbkt{\phi_S^{m\pm}}{H_0}{\phi_S^{m\pm}}$ and $\tbkt{\phi_T^{m\pm}}{H_0}{\phi_T^{m\pm}}$ are equal. As a result of this the mixed singlet and triplet subspaces always yield the same energy eigenvalues.  

\begin{figure}[t]
\includegraphics[width=0.92\columnwidth]{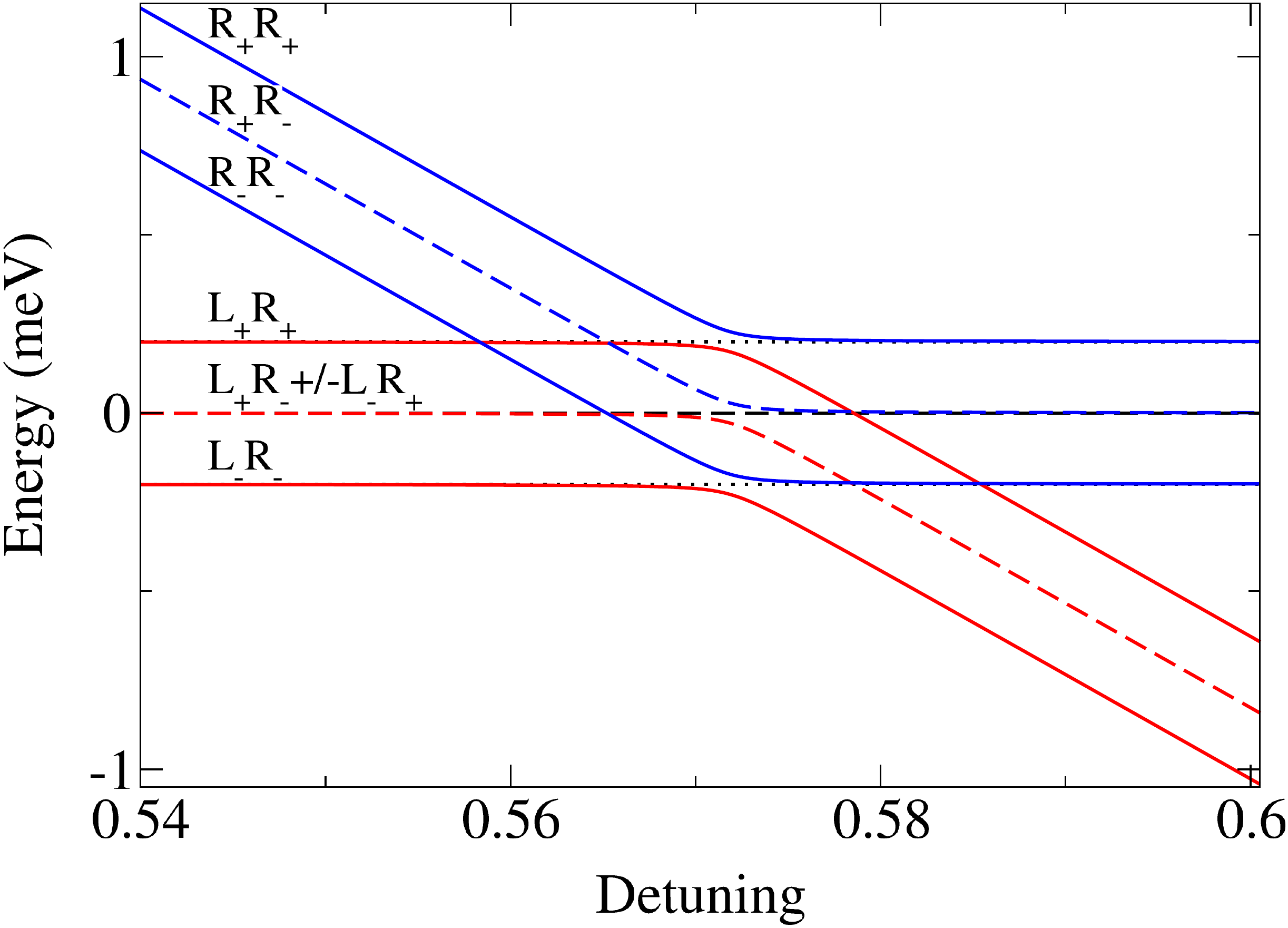}
\caption{DQD spectrum for a=8.2nm, b=3nm, d=2.45 and $\Delta$=0.1meV and zero magnetic field. The top and bottom anticrossings each consist of two singlets (solid lines) and one triplet (dotted line). In the middle anticrossing each of the three dashed lines represents a degenerate singlet/triplet level.}
\label{Anticrossing}
\end{figure}

Let us examine a concrete Si DQD with a=8.2nm, b=3nm, d=2.45 and $\Delta$=0.1meV \footnote{In this work we wish to exploit the analytical insight offered by the Hund-Mulliken approximation, which in Si/SiO$_2$ breaks down at a$\approx$10nm. We use a dot size which is somewhat unrealistic, yet not wholly unrealistic, see L.~P.~Rokhinson \textit {et al}, Phys.~Rev.~Lett. \textbf{87}, 166802 (2001).}. The energy levels are plotted in Fig.~\ref{Anticrossing} as a function of the dimensionless detuning.  At low detuning there are four (0,2) high-energy levels, indicated by the two solid lines (representing singlets of the form $\tilde{R}_+ \tilde{R}_+$, $\tilde{R}_- \tilde{R}_-$) and one dashed line (representing a degenerate singlet/triplet of the form $\tilde{R}_+\tilde{R}_-$). The separation of these levels is 2$\Delta$. There are also eight lower-energy (1,1) levels: a degenerate singlet/triplet of the form $\tilde{L}_+\tilde{R}_+$ (top solid line), a degenerate singlet/triplet of the form $\tilde{L}_-\tilde{R}_-$ (bottom solid line), and two degenerate valley-mixing singlets and triplets of the form $\tilde{L}_+ \tilde{R}_-$ and $\tilde{L}_-\tilde{R}_+$.  At high detuning the (0,2) states have lower energies than the (1,1) states. As in Ref.~[\onlinecite{PettaSci}], varying the detuning drives the energy levels towards an avoided crossing where (0,2) and (1,1) are degenerate and split by the tunnel coupling $t$ (here the splitting $\sim 6 \mu$eV).  

\begin{figure}[t]
\includegraphics[width=\columnwidth]{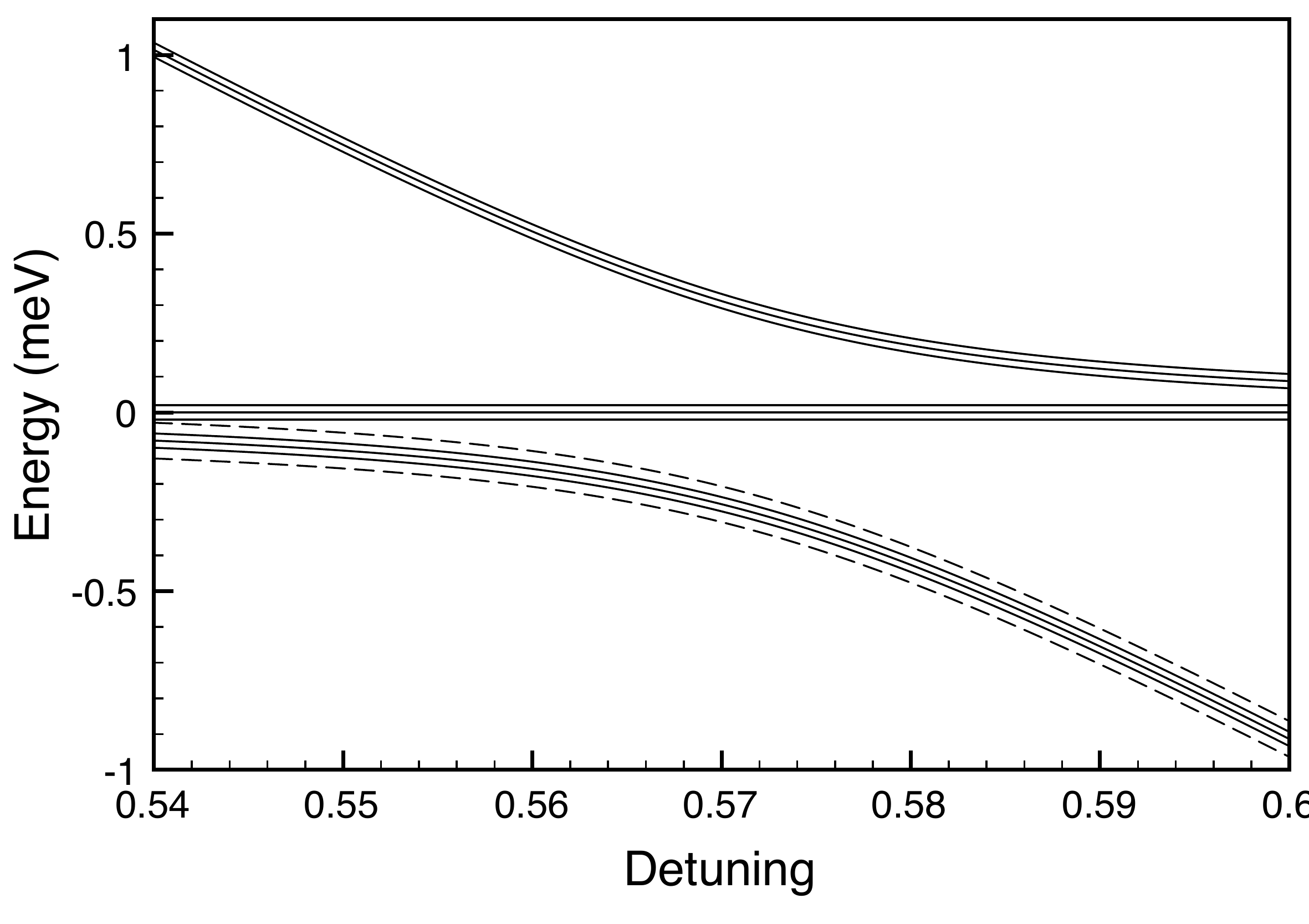}
\caption{DQD spectrum a=8.2nm, b=3nm, d=2.45, $\Delta$=0.01meV, $E_Z$ = 0.05 meV, and t=0.2meV. Dashed lines indicate Zeeman-split triplets corresponding to $T_\pm$ in Fig.~\ref{Schematic}. Other Zeeman-split levels are not shown explicitly.}
\label{SmallDelta}
\end{figure}

Thus far we have assumed that the valley splitting $\Delta$ exceeds the interdot tunnel coupling $t$. For generality, in Figure \ref{SmallDelta} we have shown the two-electron spectrum of a Si DQD when $\Delta \ll t$.  Figures.~\ref{Anticrossing} and \ref{SmallDelta} together demonstrate that, depending on the relative size of $t$, $\Delta$, and $E_Z$, the relative position of most energy levels can differ significantly, so that the loading and mixing dynamics of the two-electron states can vary dramatically. In the general case one must be prepared to expect an intermediate situation, in which a clear separation of the energy levels into three branches may not occur and some of the energy levels may cross. 

As mentioned before, in the limit in which $\Delta \gg k_B T$, the one-dot singlet state $S_1$ can be loaded exclusively.  If we now tune the bias voltage as in Ref.~[\onlinecite{PettaSci}] to shift to the (1,1) regime, the electron state will become $\phi_S^{--}$ (made from orbital states $\tilde{L}_-$ and $\tilde{R}_-$).  This state can then mix with it triplet counterpart $\phi_T^{--}$ if an inhomogeneous magnetic field is present.  No mixing with any other state is possible due to the energy separation and the fact that all inter-valley matrix elements of the Hamiltonian vanish.  When the biased voltage is tuned back to the (0,2) regime, the two electrons will either return to their initial state $\phi_S^{--}$ or stay in the $\phi_T^{--}$ triplet state and get spin-blocked.  Essentially the two-electron dynamics is confined to the lowest energy manifold of Fig.~\ref{Anticrossing} (the lowest solid curve and the lowest horizontal dotted line), in exact analogy to what happens in a GaAs DQD in Ref.~\cite{PettaSci}.  That is, pulsed manipulation and measurement can be done reliably for singlet-triplet spin qubits in a Si DQD.

\section{Measuring the valley splitting}
\label{sec:msr}
The key question for a Si DQD is whether $\Delta \gg k_B T$.  However, in general $\Delta$ is not known, so that the two-electron initialization comprises some uncertainty.  Below we explore ways to determine the valley splitting using an experiment analogous to the one described in Ref.~\cite{PettaSci}. First, we identify three loading/mixing regimes in Figs.~\ref{Schematic} and \ref{Anticrossing}.  If $S_1$ is loaded, the system is driven to the anticrossing at the bottom of Fig.~\ref{Anticrossing}, as discussed in the previous paragraph, when the detuning is varied.  If $S_2$, $T_+$, $T_0$, and $T_-$ are loaded, the system is driven to the anticrossing in the middle of Fig.~\ref{Anticrossing}.  If $S_3$ is loaded, the system is driven to the anticrossing at the top of Fig.~\ref{Anticrossing}.  One may in principle load any of the six states in Fig.~\ref{Schematic}, so that any of the three anticrossings may be involved in such an experiment. However, a magnetic field does not mix any of the three regimes of Fig.~\ref{Anticrossing}, thus once a state is loaded the experiment can be carried out as in [\onlinecite{PettaSci}] and the same readout process can be used. Interestingly, if $S_1$, $S_2$, $S_3$, or $T_0$ is loaded, the experiment is identical to Ref.~[\onlinecite{PettaSci}]. At high ${\bm B}$ (in which singlets mix only with $T_0$ triplets and vice versa) these states have an average probability of return of 1/2. If $T_+$ or $T_-$ is loaded the average probability of return will be 1 at a high magnetic field since they do not mix with other states. The average probability of return thus depends on the loading probabilities of the individual states. By studying the average probability of return one may estimate $\Delta$ by sweeping a uniform applied magnetic field (different from the inhomogeneous field mixing the singlets and triplets) as follows. 

\begin{figure}[tbp] 
\begin{tabular}{lr}
\includegraphics[width=\columnwidth]{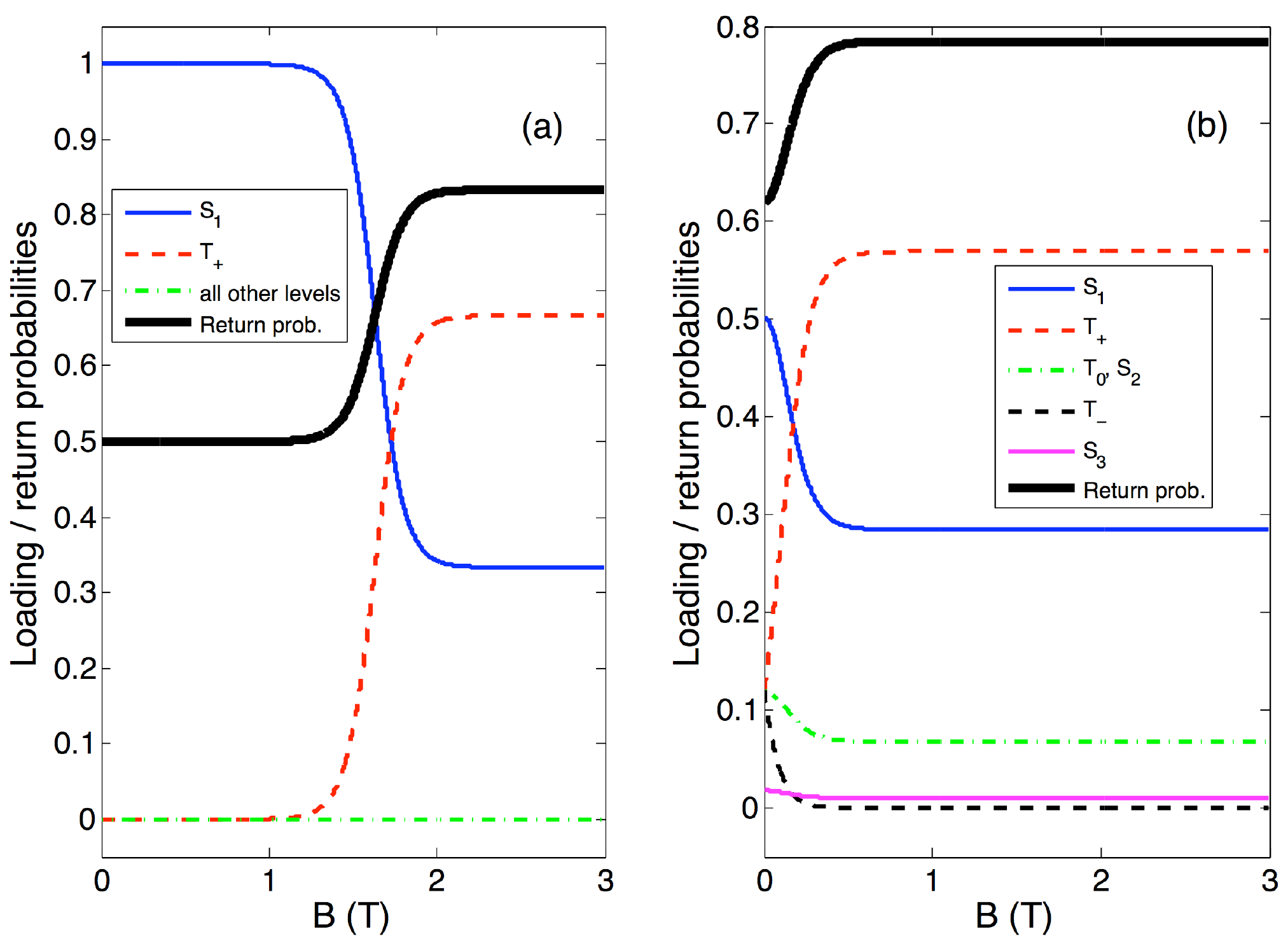}
\end{tabular}
\caption{(Color online): Loading probabilities of different levels and total return probability as a function of the magnetic field B for T=100mK and (a) $\Delta$=0.1meV and (b) $\Delta$=0.01meV.} 
\label{Loading}
\end{figure}

Sweeping a uniform magnetic field changes the loading probability of the $T_+$ triplet, which in turn leads to changes in the measurable return probability.  At low field $E_Z < 2\Delta$ and the two electrons predominantly load into the ground singlet $S_1$. Increasing the magnetic field will eventually bring $T_+$ below $S_1$, with a crossing at $E_Z = 2\Delta$.  In Fig.~\ref{Loading} we plot the loading and return probabilities against the magnetic field for $\Delta \gg$ k$_B$T in (a) and $\Delta \approx$ k$_B$T in (b).  In both cases the probability of loading $S_1$ and $T_0$ will be very close to 1/2 at the crossing point when $E_Z = 2\Delta$.  Also, the return probability increases dramatically close to the crossing point, and the crossing point corresponds approximately to where the return probability reaches the mid point between its low-field and high-field values: For $\Delta = 0.1$ meV, the crossing is at $B = 1.76$ T while the mid point is at $B = 1.63$ T; for $\Delta = 0.01$ meV, the crossing is at $B = 0.176$ T while the mid point is at $B = 0.17$ T.  The identification of the magnetic field for this mid point thus gives a reliable estimate of the value of valley splitting $2 \Delta = E_Z$.  This method should succeed as long as $2\Delta >$k$_B$T.  If $2\Delta \ll k_B T$, the return probability will not change much as we sweep the magnetic field, with the increased loading of $T_+$ compensated by the reduced loading into $T_-$.  Thus the overall change/no-change of return probability also gives a clear indication of whether $\Delta$ is larger than $k_B T$ or not.

\section{Issues specific to Silicon}
\label{sec:issues}
In GaAs the inhomogeneous magnetic field is produced by the hyperfine interaction \cite{PettaSci}. In Si the hyperfine interaction is smaller and singlet-triplet mixing will be about two orders of magnitude slower than in GaAs. Using a nanomagnet one can design a particular field magnitude and direction, enabling better control of the spin qubit. For example an inhomogeneous magnetic field along the $\hat{\bm z}$-direction mixes only the singlet and $T_0$, whereas a field along the $\hat{\bm x}$-direction mixes only the singlet and $T_\pm$. Evidently the issues discussed in this work are insensitive to the origin of the inhomogeneous magnetic field as long as this field is present in the system.

The length scale of surface roughness, which determines the spatial variation of $\Delta$, as compared with the dot size and location is not precisely known. The proposed experiments will work as long as $\Delta$ varies over a length scale larger than the DQD size, or as long as the change in $\Delta$ does not lead to change in the compositions of the valley eigenstates. We have assumed the same valley-orbit coupling in both QDs, thus the same valley splitting and eigenstates. A change in the valley composition of the eigenstates could lead to intervalley scatterings in the (0,2) to (1,1) transition, so that control of electron orbital states may become intractable. At the same time large variation of $\Delta$ across the DQD will hinder the effectiveness of the experiment. It is imperative for experimental setups to ensure firstly that the interface is as smooth as possible and secondly that the DQD spans an area over which the interface roughness profile varies as little as possible.

\section{Summary}
\label{sec:summary}
We have studied the feasibility of initialization and coherent manipulation of singlet-triplet qubits in multivalley Si DQDs, demonstrating that the valley degree of freedom makes the physics of Si quantum dots considerably different from that of dots made out of single-valley systems such as GaAs. Various experimental outcomes are possible depending on the value of the valley splitting $\Delta$. For large $\Delta$ (i.e. $\Delta \gg k_BT$) a quantum coherent experiment identical to Ref.~[\onlinecite{PettaSci}] is feasible.  For small $\Delta$ a number of different states may be initialized, leading to different experimental outcomes. One interesting highlight of our work is that, although several singlet/triplet states may be initialized, in general each state can mix with \textit{one} other state, and no more. Therefore, in principle, once a state is loaded, operations on it can proceed in a similar way to the scheme implemented in GaAs dots. For any $\Delta$, sweeping a uniform magnetic field provides a useful method for estimating $\Delta$. In fact, one very important consequence of our work is the proposed new method for estimating the valley splitting $\Delta$ in Si quantum dots, particularly when $\Delta \lesssim k_BT$. Considering the difficulties inherent in proving that a certain state belongs to a particular valley, and thus in identifying a particular energy splitting with the valley splitting, it will be important to have as many different methods as possible to measure/estimate the size of the valley splitting.

\begin{acknowledgements}
We would like to thank Neil~M.~Zimmerman, J.~M.~Taylor, Ted~Thorbeck, M.~S.~Carroll, M.~P.~Lilly, R.~P.~Muller, Erik~Nielsen, Lisa~Tracy, H.~W.~Jiang, Matt~House, M. A. Eriksson, M. Friesen, S.~N.~Coppersmith, R. Joynt, C. B. Simmons, J.~R.~Petta, Q. Niu and Zhenyu Zhang for stimulating discussions. This work is supported by LPS-NSA.  
\end{acknowledgements}

\end{document}